\documentclass[twocolumn,aps,superscriptaddress,showpacs,nofootinbib,floatfix]{revtex4}
\usepackage{epsfig,bm,feynmf}
\usepackage{graphics}

\usepackage[normalem]{ulem}  
\usepackage[dvips]{color} 
\newcommand{\com}[1]{{\sf\color[rgb]{0,0,1}{#1}}}

\renewcommand\sout{\bgroup \color{red} \ULdepth=-.5ex \ULset}

\begin{document}

\title{Contributions of hyperon-hyperon scattering to subthreshold cascade production in heavy ion collisions}

\author{Feng Li}\email{lifengphysics@gamil.com}
\affiliation{Cyclotron Institute and Department of Physics and
Astronomy, Texas A$\&$M University, College Station, TX 77843-3366,
USA}
\author{Lie-Wen Chen}\email{lwchen@sjtu.edu.cn}
\affiliation{INPAC, Department of Physics and Shanghai Key Laboratory for Particle Physics and Cosmology, Shanghai Jiao Tong University, Shanghai 200240, China}
\author{Che Ming Ko}\email{ko@comp.tamu.edu}
\affiliation{Cyclotron Institute and Department of Physics and
Astronomy, Texas A$\&$M University, College Station, TX 77843-3366, USA}
\author{Su Houng Lee}\email{suhoung@yonsei.ac.kr}
\affiliation{Institute of Physics and Applied Physics, Yonsei
University, Seoul 120-749, Korea}

\begin{abstract}
Using a gauged flavor SU(3)-invariant hadronic Lagrangian, we
calculate the cross sections for the strangeness-exchange reactions
$YY\leftrightarrow N\Xi$ ($Y=\Lambda, \Sigma$) in the Born
approximation. These cross sections are then used in the
Relativistic Vlasov-Uehling-Uhlenbeck
(RVUU) transport model to study $\Xi$ production in Ar+KCl
collisions at incident energy of 1.76A GeV and impact parameter $b=3.5$ fm.  We find
that including the contributions of hyperon-hyperon scattering
channels strongly enhances the yield of $\Xi$, leading to the
abundance ratio $\Xi^{-}/(\Lambda+\Sigma^{0})=3.38\times10^{-3}$,
which is essentially consistent with the recently measured value of
$(5.6\pm1.2_{-1.7}^{+1.8})\times 10^{-3}$  by the HADES
collaboration at GSI.
\end{abstract}


\maketitle

\section{Introduction}

The study of particle production in heavy ion collisions at energies
below their thresholds in nucleon-nucleon collisions was a topic of
extensive studies during the
1990s~\cite{aichelin,shor,gql,mosel,cassing}. The main
motivation for such study is that it offers the possibility of
extracting information on the nuclear equation of state (EOS) at
densities above that of normal nuclear matter. In particular, the
yield of strange hadrons, such as the kaon, has been
shown to be sensitive to the stiffness of the nuclear equation of
state up to three times normal nuclear matter density, with a softer
EOS giving a larger yield than a stiff EOS.
Indeed, experimental results obtained by the KaoS
Collaboration~\cite{kaos} at the Society for Heavy Ion Research (GSI) in 
Germany on the yield of kaons in heavy ion collisions at subthreshold energies
have led to the conclusion that the nuclear equation
of state at high densities is soft, consistent with an
incompressibility of about 200 MeV
extracted from the collective flow studies by the Plastic
Ball~\cite{plastic} and EOS~\cite{eos} Collaborations from Lawrence Berkeley 
Laboratory (LBL) and the E877~\cite{e877} and E895~\cite{e895} Collaborations at 
the Alternating Gradient Synchrotron (AGS) of Brookhaven National Laboratory 
(BNL). More recently, the doubly strange baryons $\Xi$ from Ar+KCl
collisions at 1.76A GeV, which is below the threshold energy of
$3.74$ GeV in a nucleon-nucleon collision, was
measured by the HADES Collaboration at GSI~\cite{HADES}. The
measured abundance ratio including the statistical and systematic
errors is $\Xi^{-}/(\Lambda+\Sigma^{0})=(5.6\pm1.2_{-1.7}^{+1.8})\times10^{-3}$.
This value is about 10-20 times larger than those given by the
statistical model~\cite{thermal} and the relativistic transport model~\cite{lw}.  
Because of the very low collision energy,
secondary reactions other than the direct reaction $NN\to N\Xi KK$
are expected to contribute significantantly to $\Xi$ production in
these collisions. In Ref.~\cite{lw}, the strangeness-exchange
reaction $\bar{K}Y\to\pi\Xi$ $(Y=\Lambda, \Sigma)$ between antikaon
and hyperon was introduced in the Vlasov-Uheling-Uhlenbeck 
(RVUU) transport model~\cite{RVUU} to study $\Xi$ production in heavy ion collisions.
The cross sections used in Ref.~\cite{lw} were taken from the
coupled-channel calculation of Ref.~\cite{chli} based on a gauged
flavor SU(3)-invariant hadronic Lagrangian.  Since there are more
hyperons than anitkaons in heavy ion collisions at this energy, the
strangeness-exchange reaction $YY\to N\Xi$ between two hyperons is
expected to be important for $\Xi$ production in these collisions.
In the present study, we use the same hadronic Lagarangian as in
Ref.~\cite{chli} to evaluate the cross sections for the reaction
$YY\to N\Xi$. For an exploratory study, these cross sections are
calculated in the Born approximation with the cutoff parameter in
the form factors at interaction vertices fitted to the cross
sections for the reactions $\bar{K}Y\to\pi\Xi$ obtained in
Ref.~\cite{chli}. For completeness, we also include the reaction
$\bar{K}N\to K\Xi$ with its cross section taken from empirically
available values.  Our results show that the inclusion of the
reaction $YY\to N\Xi$ significantly enhances the yield of $\Xi$ in
heavy ion collisions at subthreshold energies, resulting in the
abundance ratio $\Xi^{-}/(\Lambda+\Sigma^{0})=3.38\times10^{-3}$ in
Ar+KCl collisions at 1.76A GeV and impact parameter $b=3.5$ fm,
which is essentially consistent with the recently measured
experimental value. We find, however, that the contribution of the
reaction $\bar{K}N\to K\Xi$ to the $\Xi$ yield is negligible.

The paper is organized as follows. In Sec. II, we
describe the gauged flavor SU(3)-invariant hadronic Lagrangian
\cite{chli}, calculate the amplitudes for the reaction $YY \to N\Xi$
in the Born approximation, and parametrize the resulting cross
sections. In Sec. III, we introduce the
parametrization of the empirical cross section for the reaction
$\bar{K}N\to K\Xi$ as a function of the center of mass energy. We
then briefly review in Sec. IV the RVUU transport
model for high energy heavy ion collisions.
Numerical results on the time evolution of the
$\Xi$ abundance in Ar+KCl collisions at 1.76A GeV and impact
parameter $b=3.5$ fm are presented in Sec. V.
Finally, we present some discussions in Sect. VI and a summary in
Sec. VII.

\section{The hadronic model}

Possible reactions for $\Xi$ production from hyperon-hyperon
collisions are $\Lambda\Lambda\to N\Xi$, $\Lambda\Sigma\to N\Xi$,
and $\Sigma\Sigma\to N\Xi$. Cross sections for these reactions can
be evaluated using the same Lagrangian introduced in Ref~\cite{chli}
for studying $\Xi$ production from the reactions ${\bar
K}\Lambda\to\pi\Xi$ and ${\bar K}\Sigma\to\pi\Xi$. This Lagrangian
is based on the gauged SU(3) flavor symmetry
but with empirical masses. The coupling constants are taken, if
possible, from empirical information. Otherwise, the SU(3) relations
are used to relate unknown coupling constants to known ones. Also,
form factors are introduced at interaction vertices to take into
account the finite size of hadrons.

\subsection{The Lagrangian}\label{lagrangian}

As in Ref.~\cite{chli}, we use the following flavor SU(3)-invariant
hadronic Lagrangian for pseudoscalar mesons and baryons
\begin{eqnarray}\label{lag}
\mathcal{L}&=&i\,\mathtt{Tr}(\bar{B}{\partial\mkern-11 mu/}B)
+\mathtt{Tr}[(\partial_{\mu }P^{\dagger }\partial ^{\mu }P)]\nonumber\\
&+&g^{\prime }\left\{ \mathtt{Tr}\left[\left( 2\alpha -1\right)
\bar{B}\gamma ^{5}\gamma ^{\mu }B\partial _{\mu }P+\bar{B}\gamma ^{5}\gamma^{\mu }
\left( \partial _{\mu }P\right) B\right]\right\},\nonumber\\
\end{eqnarray}
where $B$ and $P$ denote, respectively, the baryon and pseudoscalar meson octets
\begin{equation}
B=\left(
\begin{array}{ccc}
\frac{\Sigma ^{0}}{\sqrt{2}}+\frac{\Lambda }{\sqrt{6}} & \Sigma ^{+} & p \\
\Sigma ^{-} & -\frac{\Sigma ^{0}}{\sqrt{2}}+\frac{\Lambda }{\sqrt{6}} & n \\
-\Xi ^{-} & \Xi ^{0} & -\sqrt{\frac{2}{3}}\Lambda
\end{array}
\right)
\end{equation}
and
{\footnotesize{
\begin{equation}
P=\frac{1}{\sqrt{2}}\left(
\begin{array}{ccc}
\frac{\pi ^{0}}{\sqrt{2}}+\frac{\eta _{8}}{\sqrt{6}}+\frac{\eta _{1}}{\sqrt{3}} & \pi ^{+} & K^{+} \\
\pi ^{-} & -\frac{\pi ^{0}}{\sqrt{2}}+\frac{\eta _{8}}{\sqrt{6}}+\frac{\eta_{1}}{\sqrt{3}} & K^{0} \\
K^{-} & \bar{K}^{0} & -\sqrt{\frac{2}{3}}\eta _{8}+\frac{\eta _{1}}{\sqrt{3}}
\end{array}\right),
\end{equation}}}
with $g^\prime$ being a coupling constant and $\alpha$ being a parameter.

For the interactions of baryons and pseudoscalar mesons with the
vector meson octet $V_\mu$,
\begin{equation}
V=\frac{1}{\sqrt{2}}\left(
\begin{array}{ccc}
\frac{\rho ^{0}}{\sqrt{2}}+\frac{\omega }{\sqrt{2}} & \rho ^{+} & K^{\ast +}\\
\rho ^{-} & -\frac{\rho ^{0}}{\sqrt{2}}+\frac{\omega }{\sqrt{2}} & K^{\ast 0}\\
K^{\ast -} & \bar{K}^{\ast 0} & \phi
\end{array}
\right),
\end{equation}
they are included by replacing the partial derivative $\partial
_{\mu }$ in Eq.(\ref{lag}) with the covariant derivative
$D_{\mu}=\partial _{\mu }-\frac{i}{2}g\left[ V_{\mu },\right]$,
where $g$ is another coupling constant.

We further include the tensor interactions between baryons and
vector mesons via the interaction Lagrangian
\begin{equation}
\mathcal{L}^{t}=\frac{g^{t}}{2m}\mathtt{Tr}\left[ \left( 2\alpha -1\right)
\bar{B}\sigma ^{\mu \nu }B\partial _{\mu }V_{\nu }+\bar{B}\sigma ^{\mu \nu}
\left( \partial _{\mu }V_{\nu }\right) B\right],
\end{equation}
with $g^t$ being the tensor coupling constant.

\subsection{Born approximation to the reactions $\Lambda\Lambda\to N\Xi$,
$\Lambda\Sigma\to N\Xi$, and $\Sigma\Sigma\to N\Xi$}

\begin{figure}[h]
\vspace{-1cm}
\centerline{\hspace{2.5cm}\includegraphics[width=0.7\textwidth]{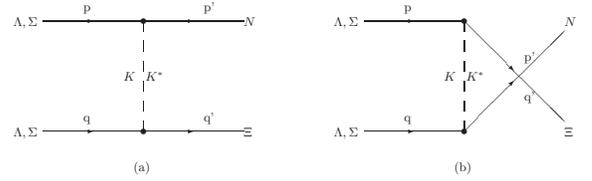}}
\vspace{-6cm} \caption{Born diagrams for the reactions
$\Lambda\Lambda\to N\Xi$, $\Lambda\Sigma\to N\Xi$, and
$\Sigma\Sigma\to N\Xi$.} \label{born}
\end{figure}

In the Born approximation, the reactions $\Lambda\Lambda\to N\Xi$,
$\Lambda\Sigma\to N\Xi$, and $\Sigma\Sigma\to N\Xi$ are described by
the tree-level $t$-channel and $u$-channel diagrams shown in Fig.~1.
To evaluate their amplitudes requires the following
interaction Lagrangian densities that are deduced from the hadronic
Lagrangian in the previous subsection, i.e.,
\begin{eqnarray}
{\cal L}_{KN\Lambda}&=&\frac{f_{KN\Lambda}}{m_K}
         \bar N \gamma^5\gamma^\mu \Lambda\partial_\mu  K+{\rm H.c.},  \nonumber\\
{\cal L}_{KN\Sigma}&=&\frac{f_{KN\Sigma}}{m_K}
         \bar N \gamma^5\gamma^\mu  (\vec\tau\cdot\vec{\Sigma}) \partial_\mu K+{\rm H.c.},  \nonumber\\
{\cal L}_{K\Lambda\Xi}&=&\frac{f_{K\Lambda\Xi}}{m_K}
    \bar\Xi\gamma^5\gamma^\mu\Lambda\partial_\mu K^c
    +{\rm H.c},\nonumber\\
{\cal L}_{K\Sigma\Xi}&=&\frac {f_{K\Sigma\Xi}}{m_K}
    \bar \Xi\gamma^5\gamma^\mu (\vec\tau \cdot\vec\Sigma)
       \partial_\mu K^c+{\rm H.c},\nonumber\\
{\cal L}_{K^*N\Lambda}&=&g_{K^*N\Lambda}\bar
    N\left(\gamma^\mu K^*_\mu+\frac{\kappa_{K^*\Lambda N}}{m_N+m_\Lambda}\sigma^{\mu\nu}
    \partial_\mu K^*_\nu\right)\Lambda\nonumber\\
    &+&{\rm H.c},\nonumber\\
{\cal L}_{K^*N\Sigma}&=&g_{K^* N\Sigma}\nonumber\\
    &\times&\vec{\bar\Sigma}\cdot\left(\gamma^\mu\vec\tau K^*_\mu+\frac{\kappa_{K^*N\Sigma}}{m_N+m_\Sigma}\sigma^{\mu\nu}\vec\tau \partial_\mu K^*_\nu\right)N\nonumber\\
     &+&{\rm H.c},\nonumber\\
{\cal L}_{K^* \Lambda \Xi }&=&g_{K^* \Lambda \Xi}\bar
    \Xi\left(\gamma^\mu K^{*c}_\mu+\frac{\kappa_{K^*\Lambda\Xi}}{m_\Lambda+m_\Xi}\sigma^{\mu\nu}\partial_\mu K^{*c}_\nu\right)\Lambda\nonumber\\
    &+&{\rm H.c},\nonumber\\
{\cal L}_{K^* \Sigma \Xi}&=&g_{K^* \Sigma \Xi}\nonumber\\
    &\times&\vec{\bar\Sigma}\cdot\left(\gamma^\mu\vec\tau K^{*c}_\mu+\frac{\kappa_{K^*\Sigma \Xi}}{m_\Sigma+m_\Xi}\sigma^{\mu\nu}\vec\tau \partial_\mu K^{*c}_\nu\right)\Xi\nonumber\\
    &+&{\rm H.c}.\nonumber\\
\end{eqnarray}
In the above, $\vec\tau$ are Pauli matrices; $\vec\pi$, $\vec\rho$,
and $\vec\Sigma$ denote the pion, rho meson, and sigma hyperon
isospin triplets, respectively; $K=(K^+,K^0)^T$
($K^*=(K^{*+},K^{*0})^T$) and $K^c=(\bar K^0,-K^-)^T$ ($K^{*c}=(\bar
K^{*0},-K^{*-})^T$) denote the pseudoscalar (vector) kaon and
antikaon isospin doublets, respectively; and $\Xi=(\Xi^0,\Xi^-)^T$
is the cascade hyperon isospin doublet. The coupling constants in
above interaction Lagrangian densities are relate to those in
Sec.~\ref{lagrangian} by
\begin{eqnarray}
\frac{f_{KN\Lambda}}{m_K}&=&\frac{2\alpha-3}{2\sqrt{3}}g^\prime,
\ \frac{f_{KN\Sigma}}{m_K}=\frac{2\alpha-1}{2}g^\prime,\nonumber\\
\frac{f_{K\Lambda\Xi}}{m_K}&=&\frac{3-4\alpha}{2\sqrt{3}} g^\prime,
\ \frac{f_{K\Sigma\Xi}}{m_K}=-\frac{1}{2}g^\prime,\nonumber\\
g_{K^*N \Lambda}&=&-g_{K^* \Lambda \Xi}=-\frac{\sqrt 3}{4}g, \nonumber\\
\ g_{K^*N\Sigma}&=&g_{K^* \Sigma\Xi}=-\frac {g}{4},\nonumber\\
\kappa_{K^*\Lambda N}&=&\frac{g^t_{K^*\Lambda N}}{g_{K^*\Lambda N}},
\ \kappa_{K^*N\Sigma}=\frac{g^t_{K^*N\Sigma}}{g_{K^*N\Sigma}},\nonumber\\
\kappa_{K^*\Lambda\Xi}&=&\frac{g^t_{K^*\Lambda\Xi}}{g_{K^*\Lambda\Xi}},
\ \kappa_{K^*\Sigma\Xi}=\frac{g^t_{K^*\Sigma\Xi}}{g_{K^*\Sigma\Xi}},\nonumber\\
\frac{g^t_{K^*N\Lambda}}{m_N+m_\Lambda}&=&\frac{2\alpha-3}{2\sqrt{3}}\frac{g^t}{2m},
\ \frac{g^t_{K^*N\Sigma}}{m_N+m_\Sigma}=\frac{2\alpha-1}{2}\frac{g^t}{2m},\nonumber\\
\frac{g^t_{K^*\Lambda\Xi}}{m_\Lambda+m_\Xi}&=&\frac{3-4\alpha}{2\sqrt{3}}\frac{g^t}{2m},
\ \frac{g^t_{K^*\Sigma\Xi}}{m_\Sigma+m_\Xi}=\com{-}\frac{g^t}{4m}.\nonumber\\
\end{eqnarray}

The cross sections for these reactions are then given by
\begin{eqnarray}
\sigma_{YY\to N\Xi}(s)=\frac{1}{64\pi sp_i^2}\int dt\overline{\left\vert \mathcal{M}\right\vert ^{2}},
\end{eqnarray}
where $s=(p_1+p_2)^2$ and $t=(p_1-p_3)^2$ are the usual squared
center of mass energy of colliding hyperons and the squared four
momentum transfer in the reaction; and $p_i$ is the momentum of 
initial hyperons in their center of mass frame.   The spin-isospin
averaged amplitude $\overline{\left\vert \mathcal{M}\right\vert
^{2}}$ in the above equation is given by
\begin{eqnarray}\label{amplitude}
&&\overline{\left\vert \mathcal{M}\right\vert ^{2}}=\frac{1}{\left( 2s_{1}+1\right) \left( 2s_{2}+1\right) \left( 2I_{1}+1\right) \left(2I_{2}+1\right)}\nonumber\\
&&\times\sum_{s_{1}s_{2}s_{1}^{\prime }s_{2}^{\prime }}
\left[ \eta _{tt}|\mathcal{M}_{s_{1}s_{2}s_{1}^{\prime }s_{2}^{\prime }}^{t}|^2
-\eta _{tu}\mathcal{M}_{s_{1}s_{2}s_{1}^{\prime }s_{2}^{\prime }}^{t}\mathcal{M}_{s_{1}s_{2}s_{1}^{\prime }s_{2}^{\prime }}^{u\ast }\right.\nonumber\\
&&-\left.\eta_{ut}\mathcal{M}_{s_{1}s_{2}s_{1}^{\prime }s_{2}^{\prime }}^{u}\mathcal{M}_{s_{1}s_{2}s_{1}^{\prime }s_{2}^{\prime }}^{t\ast }+\eta _{uu}|\mathcal{M}_{s_{1}s_{2}s_{1}^{\prime }s_{2}^{\prime }}^{u}|^2\right],
\end{eqnarray}
where  $M^t_{s_1s_2s_1^\prime s_2^\prime}$ and
$M^u_{s_1s_2s_1^\prime s_2^\prime}$ are the spin-dependent
amplitudes for the two Born diagrams shown in Fig.~\ref{born} and
are given by
\begin{eqnarray}
&&\mathcal{M}^{t}_{s_1s_2s_1^\prime s_2^\prime}(s,t)=-\frac{f_{KY_1\Xi}f_{KNY_2}}{m_K^2}
F^2({\bf p}_1-{\bf p}_3,\Lambda)\nonumber\\
&&\times\left[ \bar{\Xi}\left(p_3\right)\gamma^{5}\gamma ^{\mu }Y_1\left(p_1\right) \right]
\frac{t_{\mu }t_{\nu }}{t-m_{K}^{2}}\left[ \bar{N}\left(p_4\right) \gamma ^{5}\gamma ^{\nu}Y_2\left(p_2\right)\right]\nonumber\\
&&+g_{K^{\ast }Y_1\Xi}g_{K^{\ast }NY_2}\nonumber \\
&&\times\left[\bar{\Xi}\left(p_3\right) \left(\left( 1+\kappa _{K^{\ast }Y_1\Xi}\right) \gamma ^{\mu }-\kappa _{K^{\ast}Y_1\Xi}\frac{\left( p_3+p_1\right) ^{\mu }}{m_{Y_1}+m_{\Xi}}\right)\right.\nonumber\\
&&\times\left.Y_1\left(p_1\right)\right]\frac{g_{\mu \nu}-t_{\mu }t_{\nu }/m_{K^{\ast}}^{2}}{t-m_{K^*}^{2}}\left[\bar{N}\left(p_4\right)\left(\left(1+\kappa _{K^{\ast }NY_2}\right)\gamma^{\nu}\right.\right.\nonumber\\
&&\left.\left.+\kappa _{K^{\ast }NY_2}\frac{\left(p_3+p_1\right)^{\nu }}{m_{N}+m_{Y_2}}\right)Y_2\left(p_2\right) \right]
\end{eqnarray}
and
\begin{eqnarray}
\mathcal{M}_{s_{1}s_{2}s_{1}^{\prime }s_{2}^{\prime }}^{u}(s,u)= 
\mathcal{M}_{s_{1}s_{2}s_{1}^{\prime }s_{2}^{\prime }}^{t}(s,t),
\end{eqnarray}
with $u=(p_1-p_4)^2$. The form factor $F$ introduced at the
interaction vertex because of the hardron structure is taken to have
the monopole form,
\begin{equation}
F\left( \mathbf{q},\Lambda \right) =\frac{\Lambda ^{2}}{\Lambda ^{2}+\mathbf{q}^{2}},
\end{equation}
and depends on the three momentum transfer ${\bf q}$ and the
parameter $\Lambda$. The isospin factors $\eta_{tt}$,
$\eta_{tu}=\eta_{ut}$, and $\eta_{uu}$ in Eq.~(\ref{amplitude}),
which are obtained from summing the isospins of initial and final
particle, are 18, 10, and 18 for the reaction $\Sigma \Sigma
\rightarrow \Xi N$; 6, 2, and 6 for the reaction $\Lambda \Sigma
\rightarrow \Xi N$, and 2, 2, and 2 for the reaction $\Lambda
\Lambda \rightarrow \Xi N$.

\subsection{Cross sections for the reactions $\Lambda\Lambda\to N\Xi$,
$\Lambda\Sigma\to N\Xi$, and $\Sigma\Sigma\to N\Xi$}

\begin{table}[h]
\begin{center}
\caption{Coupling constants used in the present study.}
\vspace{0.5cm}
\begin{tabular}{ccccc}
\hline
Vertex & $f$ & Vertex & $g$ & $g^{t}$ \\ \hline
$KN\Lambda $ & -3.52 & $K^{\ast }N\Lambda $ & -5.63 & -21.5 \\
$KN\Sigma $ & 0.992 & $K^{\ast }N\Sigma $ & -3.25 & 6.31 \\
$K\Lambda\Xi$ & 0.900 & $K^{\ast }\Lambda \Xi$ & 5.63 & 6.52 \\
$K\Sigma\Xi $ & -3.54 & $K^{\ast }\Sigma\Xi $ & -3.25 & -26.4 \\
\hline
\end{tabular}
\end{center}
\end{table}

\begin{figure}[h]
\centering
\includegraphics[width=0.45\textwidth]{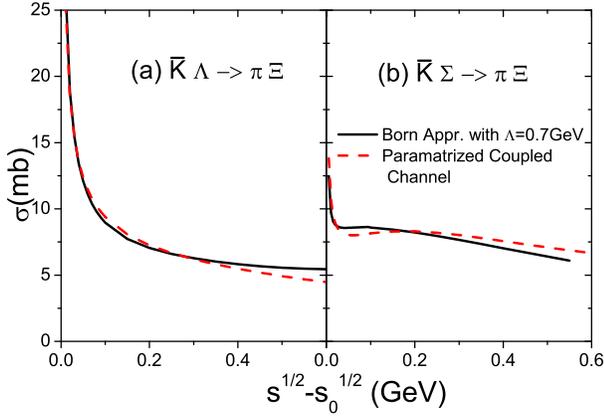}
\caption{(Color online) Isospin-averaged cross sections for  (a)
$\bar{K}\Lambda\to\pi\Xi$ and (b) $\bar{K}\Sigma\to\pi\Xi$. Solid
lines are from the Born approximation with the cutoff parameter
$\Lambda=0.7 {\rm GeV}$ in the form factor, and dashed lines are
those based on the coupled-channel calculation~\cite{chli,lw}.}
\label{ff}
\end{figure}

For numerical calculations of the cross sections, we use the
coupling constants shown in Table I. These values are obtained from
$g^\prime=14.4~{\rm GeV}^{-1}$, $g=13.0$, and $g^t/2m=19.8/m_N$ that
are determined from the empirical values $f_{\pi NN}=1.00$,
$g_{\rho NN}=3.25$, $g_{\rho NN}^{t}=19.8$ \cite{coupling_b}, and
$\alpha =0.64$~\cite{coupling_a} using relations based on the
$SU(3)$ symmetry, i.e.,
\begin{eqnarray}
\frac{f_{\pi NN}}{m_\pi}=\frac{g^\prime}{2},\ g_{\rho NN}=\frac{g}{4},\ \frac{g^t_{\rho NN}}{2m_N}=\frac{g^t}{4m}.
\end{eqnarray}
For the cutoff parameter $\Lambda$ in the form factor, its value is taken to be
$\Lambda=0.7$ GeV in order to reproduce,  as shown in Fig.~\ref{ff}, the cross sections for the reactions
$\bar{K}\Lambda \rightarrow \pi \Xi $ and $\bar{K}\Sigma \rightarrow
\pi \Xi$ that are obtained from the coupled-channel calculation based on the same hadronic
Lagrangian~\cite{chli}.

\begin{figure}
\includegraphics[width=0.45\textwidth]{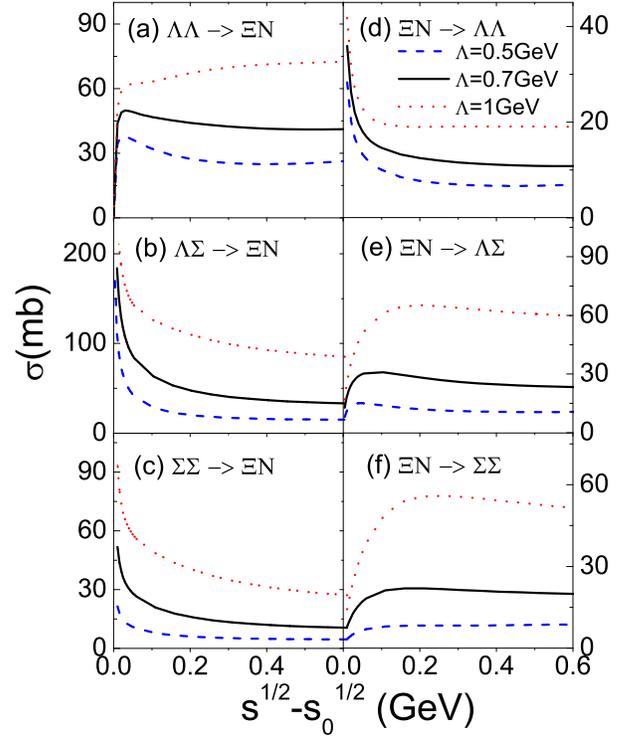}
\caption{(Color online) Cross sections for (a) $\Lambda\Lambda\to
N\Xi$, (b) $\Lambda\Sigma\to N\Xi$, (c) $\Sigma\Sigma\to N\Xi$, (d)
$N\Xi\to\Lambda\Lambda$, (e) $N\Xi\to\Lambda\Sigma$, and (f)
$N\Xi\to\Sigma\Sigma$ as functions of the center-of-mass energy
$\sqrt{s}$ from the Born approximation with cutoff parameters
$\Lambda=0.5~ {\rm GeV}$ (dashed lines), $\Lambda=0.7~{\rm GeV}$
(solid lines), and $\Lambda=1~{\rm GeV}$ (dotted lines).} \label{YY}
\end{figure}

In Fig.~\ref{YY}, we show by solid lines the isospin-averaged cross
sections for the reactions $\Lambda\Lambda\to N\Xi$ (panel (a)),
$\Lambda\Sigma\to N\Xi$ (panel (b)), and $\Sigma\Sigma\to N\Xi$
(panel (c)) as functions of the center-of-mass energy $\sqrt{s}$, obtained 
with $\Lambda=0.7~{\rm GeV}$. These cross sections can be parametrized as
\begin{eqnarray}
\sigma_{\Lambda\Lambda\to N\Xi}&=&37.15\frac{p_{N}}{p_{\Lambda}}
(\sqrt{s}-\sqrt{s_{0}})^{-0.16}~\mathrm{mb},\nonumber\\
\sigma_{\Lambda\Sigma\to N\Xi}&=&25.12(\sqrt{s}-\sqrt{s_{0}})^{-0.42} ~\mathrm{mb},\nonumber\\
\sigma_{\Sigma\Sigma\to N\Xi}&=&8.51(\sqrt{s}-\sqrt{s_{0}})^{-0.395}~ \mathrm{mb},
\end{eqnarray}
where $p_{\Lambda}$ and $p_{N}$ are initial $\Lambda$ and final nucleon momenta in the center-of-mass frame. We note that the magnitude of our cross section for the reaction $\Xi N\to\Lambda\Lambda$ is similar to that of Ref.~\cite{nakamoto98} obtained from the SU$_6$ quark model formulated in the resonance group method but is smaller than that extracted from the $(K^-,K^+)\Xi^-$ reactions in a nucleus~\cite{ahn06}. For comparisons, we also show in Fig.~\ref{YY} the cross sections for the reaction $YY\to N\Xi$ for the cutoff parameters $\Lambda=0.5~{\rm GeV}$ (dashed lines) and $\Lambda=1~{\rm GeV}$ (dotted lines). As expected, the cross sections are larger for a larger $\Lambda$.

The cross sections for the inverse reactions
$\sigma_{N\Xi\to\Lambda\Lambda}$, $\sigma_{N\Xi\to\Lambda\Sigma}$,
and $\sigma_{N\Xi\to\Sigma\Sigma}$ are related to above cross
sections by the detailed balance relations:
\begin{eqnarray}
\sigma_{N\Xi\to\Lambda\Lambda}=\frac{1}{4}\left(\frac{p_{\Lambda}}{p_{N}}\right)^{2}\sigma_{\Lambda\Lambda\to N\Xi},\nonumber\\
\sigma_{N\Xi\to\Lambda\Sigma}=\frac{3}{4}\left(\frac{p_{\Lambda}}{p_{N}}\right)^{2}\sigma_{\Lambda\Sigma\to N\Xi},\nonumber\\
\sigma_{N\Xi\to\Sigma\Sigma}=\frac{9}{4}\left(\frac{p_{\Sigma}}{p_{N}}\right)^{2}\sigma_{\Sigma\Sigma\to N\Xi}.
\end{eqnarray}

\section{Cross sections for the reaction ${\bar K}N\to K\Xi$}

\begin{figure}[h]
\includegraphics[width=0.23\textwidth]{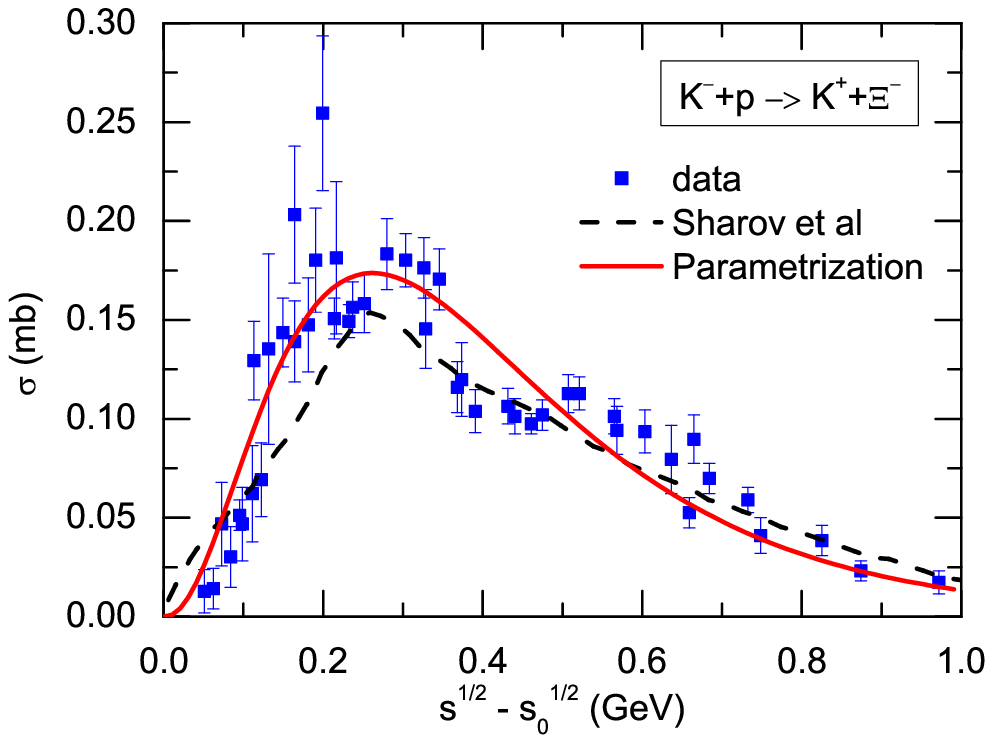}
\includegraphics[width=0.23\textwidth]{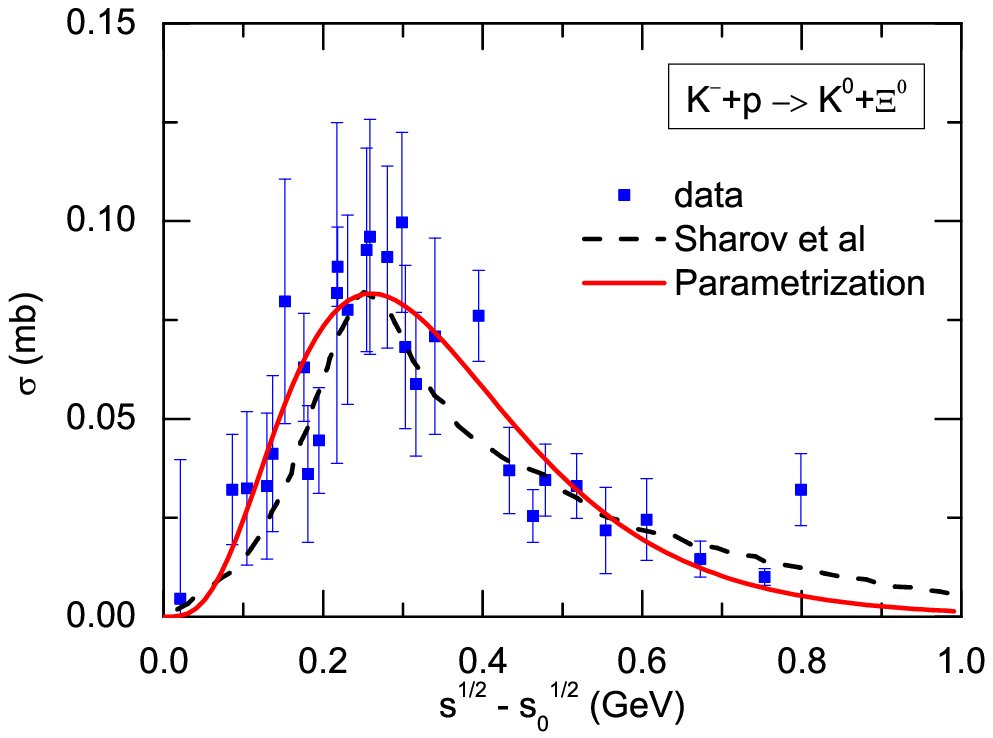}
\includegraphics[width=0.23\textwidth]{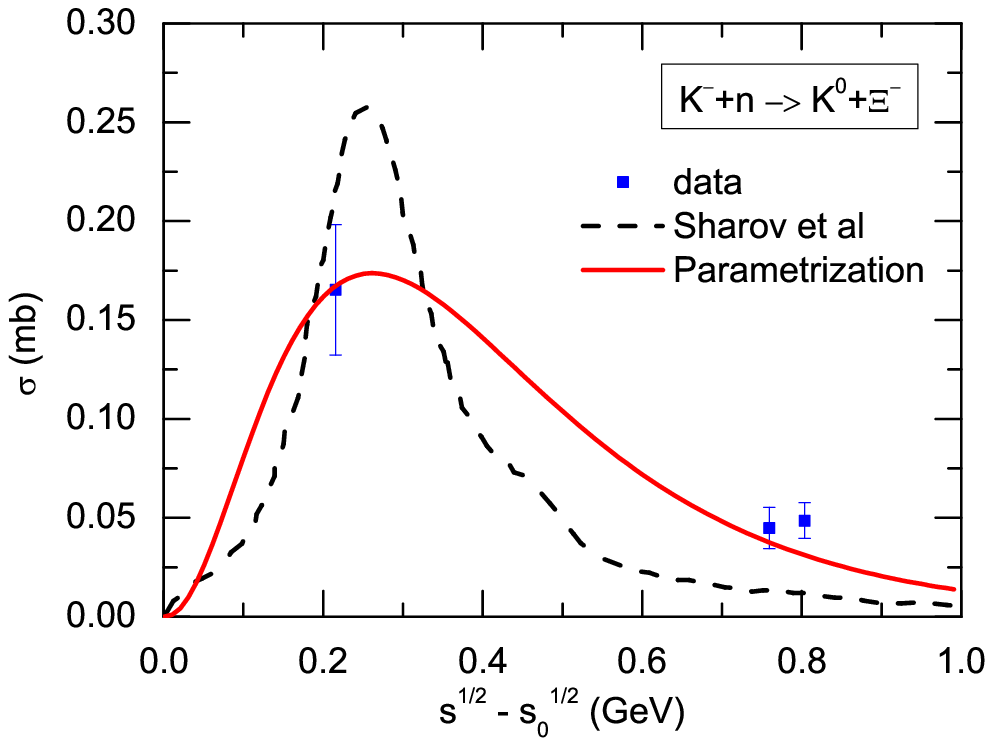}
\includegraphics[width=0.23\textwidth]{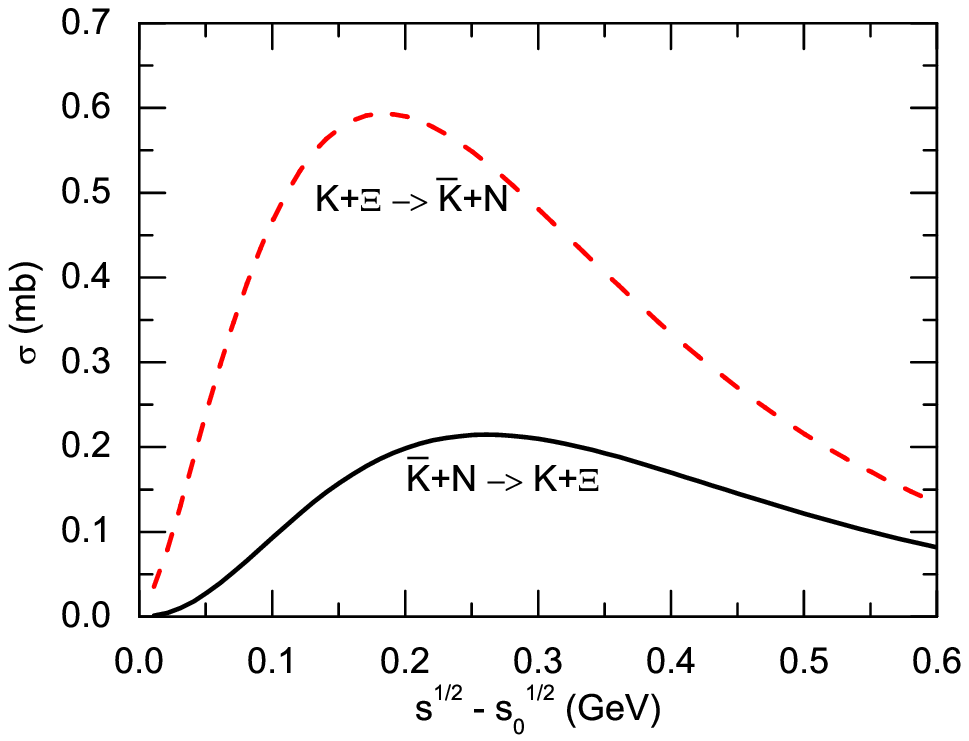}
\caption{(Color online) Cross sections for the  reaction
$K^-+p\to K^+\Xi^-$ (upper left window), $K^-+p\to K^0\Xi^0$ (upper right window), and $K^-+n\to
K^0\Xi^0$ (lower left window). Solid squares are experimental data, dashed lines are
theoretical results of Ref.~\cite{Sharov}, and solid lines are our
parametrization. Lower right window shows the isospin averaged cross sections
for the reaction ${\bar K}N\to K\Xi$ (solid line) and its inverse
reaction $K\Xi\to{\bar K}N$ (dashed line) based on our
parametrization. } \label{KN}
\end{figure}

For completeness, we also include in the present study the  reaction
$\bar{K}N\to K\Xi$.  Both the differential and total cross sections
for this reaction were measured in 1960s and 70s~\cite
{data1,data2,data3,data4,data5,data6,data7,data8,data9,data10}, and
they are shown in Fig.~\ref{KN} by solid squares for $K^-+p\to
K^+\Xi^-$ (panel (a)), $K^-+p\to K^0\Xi^0$ (panel (b)), and
$K^-+n\to K^0\Xi^0$ (panel (c)).  Recently, a phenomenological model
was introduced in Ref.~\cite{Sharov} to describe these reactions,
and the results are shown by dashed lines in Fig.~\ref{KN}. In the
present study, we use the following parametrization for these cross
sections:
\begin{eqnarray}
\sigma_{K^{-}p\to K^{+}\Xi^{-}} &=&  235.6 \left(1-\frac{\sqrt{s_{0}}}{\sqrt{s}} \right)^{2.4} \left(\frac{\sqrt{s_{0}}}{\sqrt{s}} \right)^{16.6} ~ \mathrm{mb},\nonumber\\
\sigma_{K^{-}p\to K^{0}\Xi^{0}} &=&  7739.9 \left(1-\frac{\sqrt{s_{0}}}{\sqrt{s}} \right)^{3.8} \left(\frac{\sqrt{s_{0}}}{\sqrt{s}} \right)^{26.5} ~ \mathrm{mb},\nonumber\\
\sigma_{K^{-}n\to K^{0}\Xi^{-}} &=&  235.6 \left(1-\frac{\sqrt{s_{0}}}{\sqrt{s}} \right)^{2.4} \left(\frac{\sqrt{s_{0}}}{\sqrt{s}} \right)^{16.6} ~ \mathrm{mb}.\nonumber\\
\end{eqnarray}
In terms of these cross sections, the isospin averaged cross
section for the reaction ${\bar K}N\to K\Xi$ can be expressed as
\begin{eqnarray}
\sigma_{\bar{K}N\to K\Xi}&=&0.5(\sigma_{K^{-}p\to K^{+}\Xi^{-}}+\sigma_{K^{-}p\to K^{0}\Xi^{0}}\nonumber\\
&+&\sigma_{K^{-}n\to K^{0}\Xi^{-}}).
\end{eqnarray}
The detailed balance relation then allows us to express the cross section for
the inverse reaction $K\Xi\to{\bar K}N$ as
\begin{equation}
\sigma_{K\Xi\to\bar{K}N}=\left(\frac{p_{N}}{p_{\Xi}}\right)^{2}\sigma_{\bar{K}N\to K\Xi}
\end{equation}
where $p_N$ and $p_\Xi$ are the 3-momenta of  nucleon and $\Xi$ in
the center-of-mass frame.

\section{the relativistic Vlasov-Uhling-Ulenbeck transport model}

To study $\Xi$ production in heavy ion collisions at subthreshold energies,
we generalize the RVUU transport model~\cite{RVUU} to include the reactions $YY\to N\Xi$
and ${\bar K}N\to K\Xi$ and their inverse reactions besides the
reaction ${\bar K}Y\to \pi\Xi$ and its inverse reaction that were
already included in Ref.~\cite{lw}. In addition to  these reactions
and other reactions involving nucleons, Delta resonances, hyperons,
pions, kaons, and antikaons, the VUU model also includes the
mean-field effect on the propagation of baryons, kaons, and
antikaons. For nucleons and Delta resonances,  their mean-field
potentials are taken from the relativistic mean-field model via the
scalar and vector potentials, so their motions are given by the
following equations of motion:
\begin{eqnarray}
\dot{\mathbf{x}}&=&\frac{\mathbf{p}^{*}}{E^{*}}\nonumber\\
\dot{\mathbf{p}}&=&-\nabla_{x}(E^{*}+W_{0})
\end{eqnarray}
where $m^{*}=m-\Phi$, $\mathbf{p}^{*}=\mathbf{p}-\mathbf{W}$,
$E^{*}=\sqrt{\mathbf{\mathrm{p}}^{*2}+m^{*2}}$ with $\Phi$ and
$W=(W_{0},\mathbf{W})$ being the scalar and vector mean fields,
respectively. These mean fields are calculated from the effective
chiral Lagrangian of Ref.~\cite{chiral} with parameters determined
from fitting the nuclear matter incompressibility
$K_{0}=194\mathrm{MeV}$ and the nucleon effective mass $m^{*}/m=0.6$
at normal nuclear matter density $\rho_{0}=0.15~\mathrm{fm^{3}}$.
For $\Lambda$ and $\Sigma$ hyperons, their mean-field potentials are
taken to be 2/3 of the nucleon mean-field potential according to
their light quark content. Similarly, the mean-field potential for
$\Xi$ is 1/3 of that of the nucleon .

For kaons and antikaons, their mean-field potentials are derived,
on the other hand, from the dispersion relation obtained in the
chiral Lagrangian~\cite{gqli}
\begin{equation}
 \omega_{K,\bar{K}}=\left [m_{K,\bar{K}}^{2}+\mathbf{p}^{2}-\frac{\Sigma_{NK}}{f^{2}}\rho_{s}+\left (\frac{3}{8}\frac{\rho_{N}}{f^{2}} \right )^{2} \right ]^{1/2}\pm\frac{3}{8}\frac{\rho_{N}}{f^{2}},
 \label{Ek}
\end{equation}
where $\rho_{s}=\langle\bar{N}N\rangle$ is the scalar density,
$f=103$ MeV is the pion decay constant, and the $\pm$ is taken as
\lq\lq{+}\rq\rq{} for kaons and \lq\lq{-}\rq\rq{} for antikaons.
The $KN$ and $\bar{K}N$ sigma term $\Sigma_{NK}$ in the above
equation can in principle be calculated from the $SU(3)_{L}\times
SU(3)_{R}$ chiral Lagrangian but are taken to have the values
$\Sigma_{NK}/f^{2}=0.22~ \mathrm{ GeV^{2}fm^{3}}$ and
$\Sigma_{N\bar{K}}/f^{2}=0.35~ \mathrm{ GeV^{2}fm^{3}}$ as in
Ref.\cite{lw} from fitting the kaon and antikaon yields in heavy ion
collisions.

Besides affecting the propagation of particles, the mean-field
potential also has effect on the threshold energy for particle
production as a result of the potential difference between the
initial and final states of a reaction. For example, this effect is
important for understanding the enhanced production of antikaon
through the reactions $BB\leftrightarrow BBK\bar{K}$, $\pi
B\leftrightarrow K\bar{K}B$, and $\pi Y\leftrightarrow\bar{K}N$ in
heavy ion collisions at subthresold energies. As a result, the
contribution of the reaction ${\bar K}Y\to \pi\Xi$ to $\Xi$
production in heavy ion collisions at subthreshold energies was
found in Ref.~\cite{lw} to be further enhanced.  We note that in the
RVUU model, kaons, antikaons, hyperons (lambdas and sigmas), and cascade particles
are treated perturbatively by neglecting the effect of their production and annihilation on the collision dynamics, 
which is dominated by the more abundant nucleons, Delta resonances, and pions.
In this approach, kaons, antikaons, and hyperson are produced from nucleon (Delta)-nucleon (Delta) 
and pion-nucleon (Delta) collisions whenever it is energetically allowed, and they are given probabilities 
that are determined by the ratios of their respective production cross sections to the total cross sections of the colliding 
particles.  For $\Xi$ production from antikaon collisions with nucleons or hyperons
and from hyperon-hyperon collisions, it is similarly treated but the probability of the produced $\Xi$ 
is reduced by the probabilities of colliding particles. The annihilation of these rare particles
is treated in a similar way and leads to reductions of their probabilities.  
The present approach thus takes into account the small probability associated with the
production of two rare particles in a subthrehold heavy ion collision that are involved in the production of a $\Xi$.

\section{results}

\begin{figure}[h]
\includegraphics[width=0.45\textwidth]{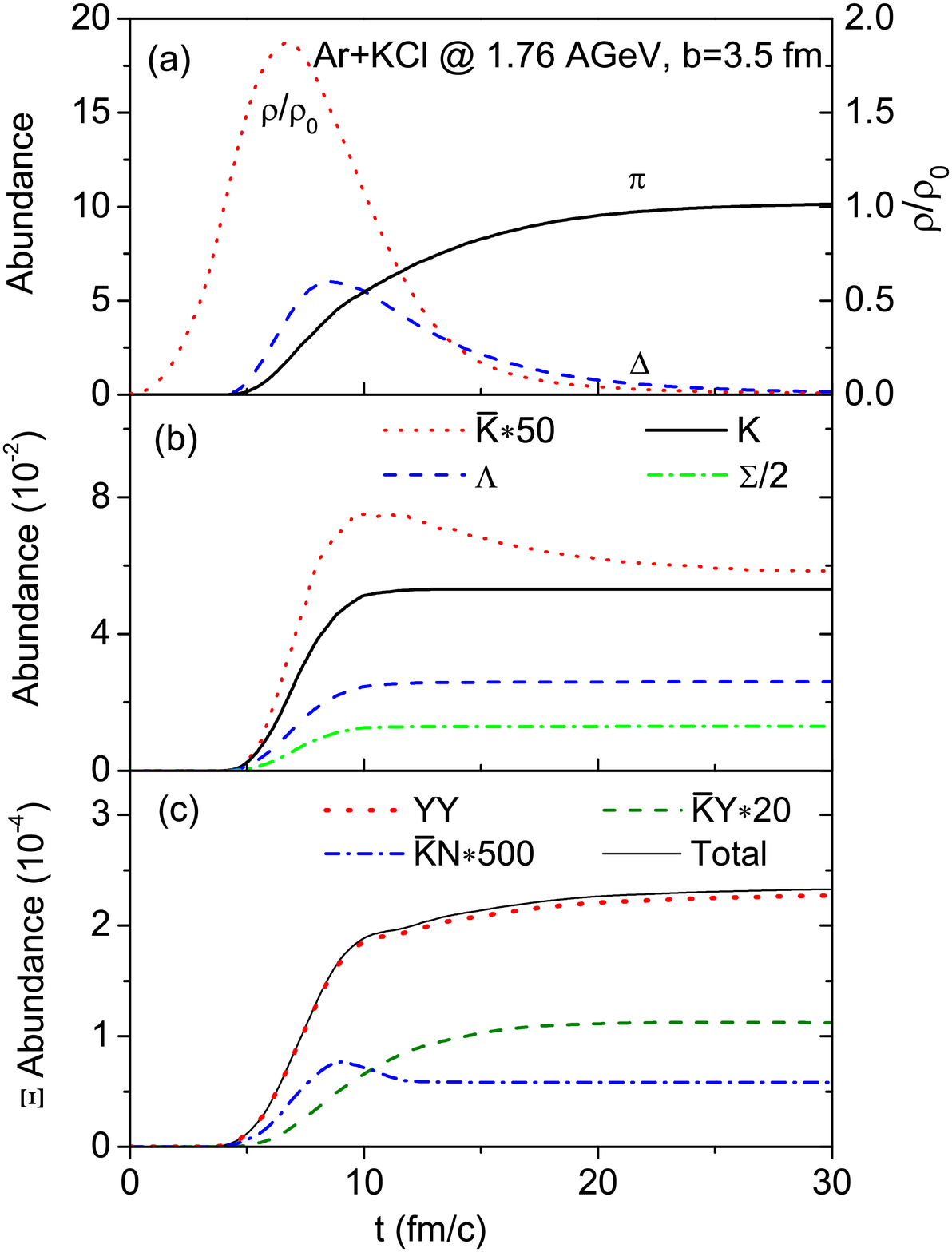}
\caption{(Color online) Time evolutions of (a) central baryon
density (right scale) and the abundances (left scales) of
$\pi$, $\Delta$; (b) $K$, $\Lambda$, $\Sigma$, and $\bar{K}$; and (c)
$\Xi$ produced from different reactions.}\label{Abundance}
\end{figure}

In this Section, we show the results for $^{40}\mathrm{Ar}+\mathrm{KCl}$
collisions at incident energy 1.76 AGeV, taking as an
average of $^{40}\mathrm{Ar}+\mathrm{K}^{39}$ collisions and
$^{40}\mathrm{Ar}+\mathrm{Cl}^{35}$
collisions, and compare them with the data from the HADES
Collaboration at SIS. The HADES trigger (LVL1) selects approximately
the most central $35\%$ of the total reaction cross
section~\cite{HADES}. According to GEANT simulations \cite{GEANT}
with the UrQMD \cite{QMD1,QMD2} transport approach as event
generator, the average value and width of the corresponding impact
parameter distribution amount to 3.5 and 1.5 fm, respectively. For
simplicity, we take $b=3.5$ fm in the present study.
Fig.~\ref{Abundance}(a) shows the time evolution of $\pi$ and
$\Delta$ abundances (left scale) and the central baryon density
(right scale). It is seen that the colliding system reaches its
highest density of about $1.87\rho_{0}$ at about 7 $\mathrm{fm}/c$
when most particles are produced. The $\pi$ abundance saturates at
$10.3$.  Assuming isospin symmetry, the $\pi^-$ number is then 3.43 which is
very close to the measured number of $3.9\pm 0.1\pm 0.1$ by the HADES 
Collaboration~\cite{HADES2,HADES3}. 
The time evolution for the abundances of $K$, $\bar{K}$, $\Lambda$, and $\Sigma$ 
are shown in Fig.~\ref{Abundance}(b), and they saturate at the values of
$5.32\times10^{-2}$, $1.15\times10^{-3}$, $2.60\times10^{-2}$, and
$2.60\times10^{-2}$, respectively.  Assuming isospin symmetry gives
$2.61\times10^{-2}$ for the $K^+$ number, $5.75\times10^{-4}$ for the 
$K^-$ number, and $3.47\times10^{-2}$ for the $\Lambda+\Sigma^0$ number.
These numbers are again close to corresponding measured numbers of 
$(2.8\pm 0.2\pm 0.1\pm 0.1)\times 10^{-2}$, $(7.1\pm 1.5\pm 0.3\pm 0.1)\times 10^{-4}$,  
and $(4.09\pm 0.1\pm 0.17)\times 10^{-2}$ by the HADES Collaboration~\cite{HADES4}.
For the time evolution of the $\Xi$ abundance, it is shown by the solid curve in
Fig.\ref{Abundance}(c) and is seen to saturate at the value
$2.34\times10^{-4}$.   Taking $\Xi^{-}$ as half of $\Xi$ by assuming isospin symmetry,
we obtained a $\Xi^-$ number of $1.17\times 10^{-4}$ which is about half of the measured
number of $(2.3\pm 0.9)\times 10^{-4}$ by the HADES Collaboration~\cite{HADES5}.  
Our results thus lead to an abundance ratio $\Xi^{-}/(\Lambda+\Sigma^{0})=3.38\times10^{-3}$, 
which is essentially consistent with the measured value of $(5.6\pm1.2_{-1.7}^{+1.8})\times 10^{-3}$ by the HADES collaboration.

The contributions to $\Xi$ production from different reaction
channels are also shown in Fig.\ref{Abundance}(c). Dotted,
dashed-dotted, and dash lines denote, respectively, the abundance of
the $\Xi$ particles from the reactions $YY\to N\Xi$,
$\bar{K}Y\to\pi\Xi$, and $\bar{K}N\to K\Xi$.  Compared to the total
$\Xi$ abundance, shown by the solid line in Fig.\ref{Abundance}, the
contributions are $97.5\%$, $2.40\%$, and $0.1\%$ from the reactions
$YY\to N\Xi$, $\bar{K}Y\to\pi\Xi$, and $\bar{K}N\to K\Xi$,
respectively. So the $YY\to N\Xi$ channel dominates $\Xi$ production
in heavy ion collisions at subthreshold energies. This can be
explained by the fact that the cross section for $YY\to N\Xi$ is
almost 3-4 times the cross section for $\bar{K}Y\to\pi\Xi$, and
almost hundred times the cross section for $\bar{K}N\to K\Xi$. Also,
the hyperon abundance in the system is almost 20 times the anti-kaon
abundance. We note that the relative contributions to the $\Xi$ yield from the reactions
$\Lambda\Lambda\to N\Xi$, $\Lambda\Sigma\to N\Xi$ and $\Sigma\Sigma\to N\Xi$
are about 1, 4 and 1. 

\section{discussions}

Our results are obtained without the consideration of the
isospin asymmetry effect due to different proton and neutron numbers
in the colliding nuclei, which is expected to increase the final
abundance ratio $\Xi^{-}/(\Lambda+\Sigma^{0})$. If we assume that
the abundance of $\Xi$ has reached chemical equilibrium in heavy ion collisions, 
which is certainly questionable in view of the failure of the statistical model 
in describing the experimental data, this enhancement 
can be estimated using $\Xi^{-}/\Xi^{0}=\mathrm{e}^{-\mu_{C}/T}=\Sigma^{-}/\Sigma^{0}=\Sigma^{0}/\Sigma^{+}=N/Z$, where $\mu_{C}$ is the charge chemical potential and $T$ is the
temperature of the system.  With the value $N/Z\sim1.14$ for $\mathrm{Ar}^{40}+\mathrm{K}^{39}$ or $\mathrm{Ar}^{40}+\mathrm{Cl}^{35}$, we have $\Xi^{-}=0.533~\Xi$ and
$\Sigma^{0}=0.3314~\Sigma$, leading to the ratio
$\Xi^{-}/(\Lambda+\Sigma^{0})=3.60\times10^{-3}$ that is $6.5\%$
larger than that for an isospin symmetric system.

Also, the nuclear EOS used in the transport model can affect the
final $\Xi$ abundance in heavy ion collisions. The results presented
in the previous Section are based on a soft EOS. Using a
stiff  EOS, we find that the $\Lambda$, $\Sigma$,
and $\Xi$ abundances are reduced to $1.74\times10^{-2}$,
$1.77\times10^{-2}$, and $1.46\times10^{-4}$, respectively. The
reason for this reduction in the hyperon abundances is that the
energy density of the colliding system increases faster for a stiff
EOS, thus making its  expansion faster and reaction time short.
However, the abundance ratio
$\Xi^{-}/(\Lambda+\Sigma^{0})=3.13\times10^{-3}$ for the stiff EOS
is essentially the same as that for a soft EOS.

Furthermore, the results presented here are for the impact parameter
$b=3.5$ fm. A more realistic comparison with experimental data
should include a distribution of impact parameters. We have checked
that using different impact parameters, the ratio
$\Xi^{-}/(\Lambda+\Sigma^{0})$ remains, however, essentially
unchanged, since both hyperons and cascade abundances change by
almost the same factor.

Finally, because of the very large $\Xi$ production cross
sections and the small size of the colliding system, the geometrical
treatment of $\Xi$ production from hyperon-hyperon scattering in
terms of their scattering cross section as used in the RVUU
transport model may become inaccurate. This can be seen from the
dependence of final $\Xi$ abundance on the value of the cutoff
parameter $\Lambda$ in the form factor used in evaluating the cross
sections of the reactions $YY\to N\Xi$. As shown in Fig.~\ref{YY},
these cross sections increase with increasing value of $\Lambda$.
Results from our transport model study show, on the other hand, that the $\Xi$ abundance
increases with decreasing value of $\Lambda$.  However, our conclusion in the present work is expected to remain unchanged since the $\Xi$ abundance changes only by about $30\%$ when the $\Xi$ production cross sections change by more than a factor of 4.  We note that a more accurate treatment of particle scattering may be achieved by using the stochastic method of Ref.~\cite{cassing02} based on the transition probability, and we hope to purse such an improved study in the future.  

\section{Summary}

We have calculated the cross sections for the reaction $YY\to N\Xi$
($Y=\Lambda$, $\Sigma$) based on a gauged SU(3)-invariant hadronic
Lagrangian in the Born approximation and found that these cross sections are
almost 4 times the cross sections for the reaction $\bar{K}Y\to\pi\Xi$ that was considered in 
previous studies.   We then used these cross sections to study $\Xi$ production in 
$^{40}\mathrm{Ar}+\mathrm{KCl}$ collisions at the subthreshold energy of 1.76 AGeV 
within the frame work of a relativistic transport model that includes explicitly the nucleon, $\Delta$, 
pion, and perturbatively the kaon, antikaon, hyperons and $\Xi$. We found that the reaction 
$YY \to N\Xi$ would enhance the $\Xi$ abundance by a factor of about 16 compared to that from the reaction ${\bar K}Y\to\pi\Xi$, resulting in abundance ratio $\Xi^{-}/(\Lambda+\Sigma^{0})=3.38\times10^{-3}$ that is essentially consistent with that measured by the HADES Collaboration at GSI. Our study has thus helped in resolving one of the puzzles in particle production from heavy ion collisions at subthrehold energies. 

\section*{Acknowledgements}

This work was supported in part by the U.S. National Science
Foundation under Grant Nos. PHY-0758115 and PHY-1068572, the Welch
Foundation under Grant No. A-1358, the NNSF of China under Grant
Nos. 10975097 and 11135011, the Shanghai Rising-Star Program
under grant No. 11QH1401100, the "Shu Guang" project
supported by Shanghai Municipal Education Commission and Shanghai
Education Development Foundation, the Program for Professor of
Special Appointment (Eastern Scholar) at Shanghai Institutions of
Higher Learning, the Science and Technology Commission of Shanghai
Municipality (11DZ2260700), and the Korean Research Foundation
under Grant No. KRF-2011-0020333.

\end{document}